\documentclass[conference]{IEEEtran}
\usepackage{booktabs}
\usepackage{url}

\usepackage{graphicx}
\usepackage{algorithm}
\usepackage{algorithmic}
\usepackage[usenames,dvipsnames]{xcolor}
\usepackage{subfigure}

\begin{document}

\title{Deploying an Information Centric Smart Lighting System in the Wild}

\author{\IEEEauthorblockN{Upeka De Silva$^{1}$, Adisorn Lertsinsrubtavee$^{2}$, \\ Arjuna Sathiaseelan$^{2}$, Carlos Molina-Jimenez$^{2}$, Kanchana Kanchanasut$^{1}$ }\\
\IEEEauthorblockA{$^{1}$Asian Institute of Technology $^{2}$University of Cambridge\\
Email: st116384@ait.asia, al773@cam.ac.uk, as2330@cam.ac.uk,  cm770@cam.ac.uk, kanchana@ait.asia}}

\maketitle
\begin{abstract}
In this paper, we present a NDN based smart home lighting solution where lights are automatically controlled in near real time based on occupancy and daylight. 
We implemented a reliable solution using NDN architecture exploiting the primitive NDN features in push based data dissemination, multicast forwarding through name prefixes and Interest filtering in application layer. 
Performance was evaluated benchmarking with respect to a cloud based approach in terms of message delivery latency. Scalability of the solution was also analyzed presenting an analysis on FIB scalability based on Interest filtering. Finally, from this study, we recommend and highlight a few requirements that could improve NDN for sustainable IoT applications. 

\end{abstract}

\vspace{-2.5ex}
\section{Introduction}
The Internet of Things (IoT) is now gaining significant momentum with efforts by the scientific community and industry to evolve the current silos based IoT platform models towards a globally unified IoT platform - a platform that needs to efficiently support 50-100 billion networked objects. This has not been a trivial task with several research and standardisation efforts proposing several solutions to build a unified host centric IoT platform as an overlay on top of today's host-centric Internet. However, several challenges still remain~\cite{zhang-icnrg-icniot}. 

In enabling this rather ambitious and challenging effort, the networking research community has turned to the Information Centric Networking (ICN) paradigm as a potential architectural solution to solve the challenges posed by IoT. ICN offers the following built-in abstractions: it uses name based routing to deliver packets
and provides inherent multicast. Likewise, it offers developers a great flexibility over naming and security. It avoids dependencies on separate protocols and various middleware in IoT networks. There are several realisations of the ICN paradigm that to a greater or lesser extend implement the abstractions mentioned above and needed
in this discussion. In this paper we will use the NDN (Named Data Networking) because it is one of the most mature realisations.

In this paper we argue that smart home ecosystems can be naturally implemented using the ICN paradigm and discuss the research challenged that need to be
addressed. To support our arguments, we conduct a feasibility study of innovative smart home solutions that can be implemented using the NDN architecture. We identified applications that are possible to realize with low cost hardware and software; we analyze their communication requirements. In summary, the key contributions 
of this paper are the following:

\begin{itemize}

	\item{We demonstrate how a smart lighting system for controlling the lights of a room that takes into consideration several parameters such as
daylight and occupancy can be implemented on the basis of the NDN architecture. A salient feature of our systems is that it can be implemented using off--the--shelf low-cost devices (e.g., cheap sensor, actuator and normal home light bulb).}

	\item{We show that the system can meet time constrains that are typical of smart home applications such as response in real time to certain events like presence
of people in a room. To meet that challenge, we selected appropriate communication models considering latency as well as message overhead and used Interest retransmission to ensure reliability considering application requirements.}

	\item{We show that the implementation is scalable. We evaluated the scalability of the network while extending our
solution to an exhibition hall scenario. Our method using Interest filtering can efficiently reduce the FIB (Forwarding Information Base) sizes of forwarding nodes compared to the 
primitive NDN.}
  		\item{Through our implementation and deployment, we demonstrate that features supporting efficient multi source communication and push based communication are vital to deploy NDN for IoT requirements.}
\end{itemize}

\vspace{-1ex}
\section{Related work}
The concept of named data networking things is recently proposed to provide a single networking solution for IoT communication instead of relying on a set of protocols or some middlewares~\cite{shang:IoTDI_16}. 
To explore this revolutionary vision, Amadeo et al.~\cite{marica:NDN_IoT_archi} present a high-level architecture for NDN IoT systems indicating key challenges for the new communication protocol. The same authors also proposed a theoretical analysis on a NDN based smart home system~\cite{amadeo:ICC_home}. Biswas et al. develop a contextualized information centric home network which has been demonstrated through the CCNx based homenet platform~\cite{Biswas:2013, Ravindran:2013} . However, this work only focused on the monitoring scenario.
With regards to the large scale NDN-IoT implementations, the NDN based sensor monitoring system and a lighting control system have been developed as a part of university campus~\cite{Burke:NCA-2014,Burke:NOMEN13}. In the lighting control system, they have focused on architectural lighting requirements, controlling lights based on patterns for a theatrical environment while the controlling of lights based on sensing has been remained for future work ~\cite{Jeff:NDN_Tech}.
Another interesting issue in developing ICN-IoT system refers to the routing protocols. However, there is no particular NDN routing protocol has been developed for IoT networks, while the flooding of Interests, called Vanilla Interest Forwarding (VIF) is considered as the common feasible approach~\cite{marica:NDN_WSN, Emmanuel:NDN_in_the_wild}. 

Our research demonstrates a use case of smart lighting system under the NDN architecture, including both sensing and actuating. Furthermore, we aim to leverage NDN inherent multicast forwarding feature and interest filtering to support effective group communication. 

\section{NDN Smart Lighting Solution}
In our solution, smart lighting control is operated automatically based on the occupancy and daylight, together with user preferences. 
For simplicity, controlling is defined as switching ON/OFF lights. We designed an efficient smart lighting solution under NDN architecture to control lights in near real time, based on the conditions derived from sensor data.

	\subsection{NDN overview}
	NDN implements a simple request/response architecture based on a stateful forwarding plane using two types of packets named \emph{Interest} and \emph{Data}, both carrying URI-like names. 

Each node in the network acts as a NDN router and maintains three types of data structures: Forwarding Information Base (FIB), Pending Interest Table (PIT), and Content Store (CS). 
FIB records matching outgoing interface(s) for particular name prefixes. 

To fetch content, a consumer sends an Interest packet into the network containing name of the required content. When a NDN node receives an Interest message, it first queries matching data in its local CS.  If the data is available,  the matching Data is sent back to the consumer through the same interface. 

Otherwise the node updates the PIT table with the Interest's name and the incoming interface. 
In case no existing PIT record is found, the node forwards the Interest over the recorded outgoing interface(s) in the FIB~\cite{NFD_developper}. 

When the Interest reaches a potential data provider or a node having a matching Data in its cache, a Data packet is always generated and replied back to the consumer, following the chain of the intermediate nodes. During the forwarding process, each node replicates Data packet to all recorded incoming interfaces in the matching PIT entry, keeps a copy in local CS and then deletes the related PIT record. 
Thus the traffic in NDN is self regulated, since it maintains at most one Data packet per one Interest in each link.

\subsection{System Architecture}
\begin{figure}[!ht]
  \centering
    \includegraphics[width=0.45\textwidth]{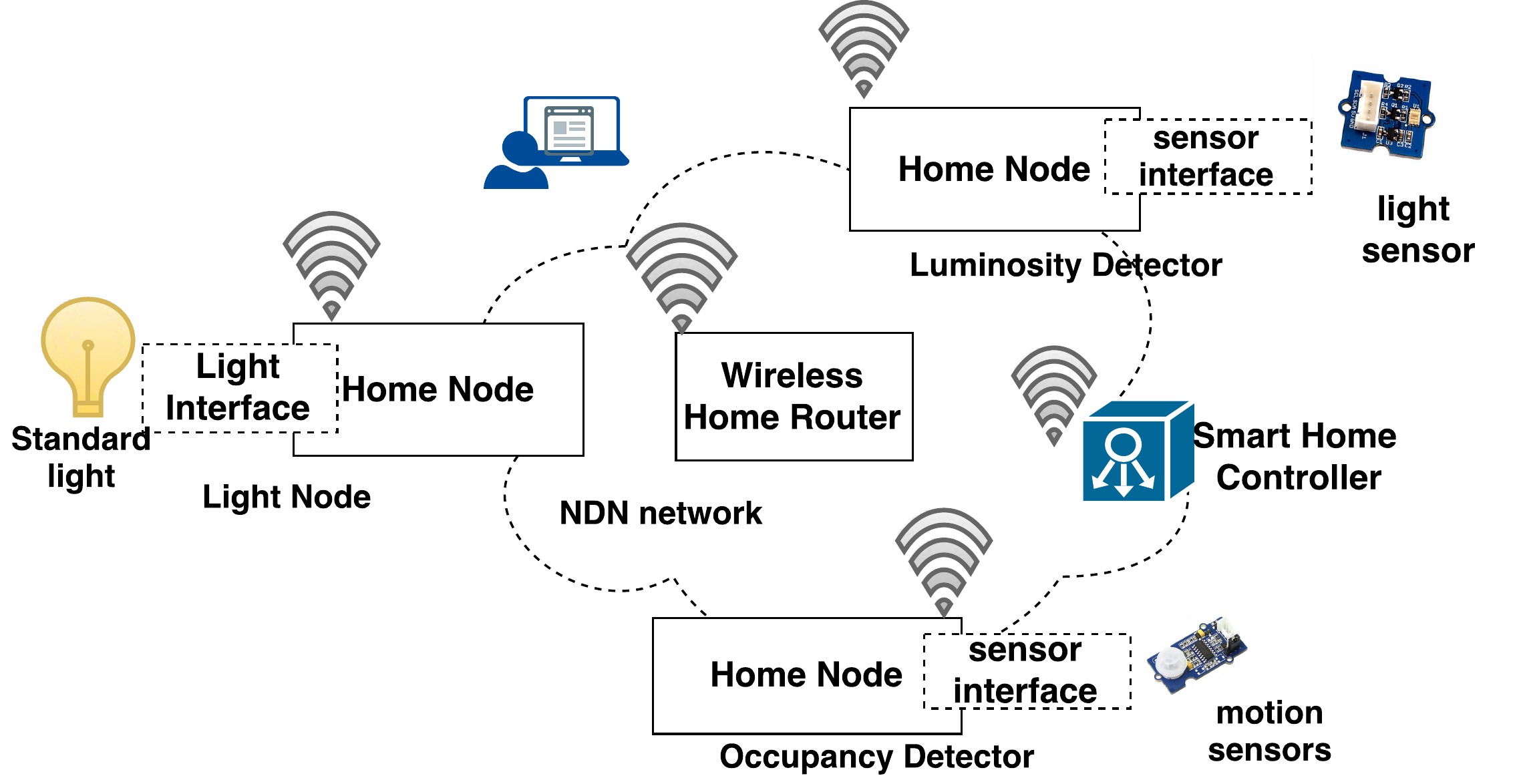}
     \caption{NDN smart lighting architecture}
      \label{architecture}
\end{figure}

Our NDN smart lighting system architecture is shown in Figure~\ref{architecture} and it is comprised of the following components:
\emph{Home Router:} A wireless router that connects all home nodes. 
\emph{Home Node:} A Raspberry Pi with a wireless interface (802.11g). 
\emph{Light Node:} A conventional light bulb, connected to a Raspberry Pi via an 
actuator that can switch the light bulb ON/OFF remotely.
\emph{Occupancy Detector:} A motion sensor circuit that tracks users' in/out movements and to/from the room. 
\emph{Luminosity Detector:} A light sensor for measuring the light level in lux unit.
\emph{Smart Home Controller:} A smart home application running on a Raspberry Pi or on a commodity laptop to control lights automatically.

\vspace{-1ex}

Raspberry Pis and the home router were installed with Raspbian Wheezy and OpenWrt 12.09 respectively. NDN Forwarding Daemon\footnote{\scriptsize{http://named-data.net/doc/NFD/current/}} (NFD) binaries were used to enable NDN functionalities. End applications were developed using python and NDN supportive client library (PyNDN). Our sensor devices (e.g., light sensors, PIR motion sensor) rely on the Grove tool kit\footnote{\scriptsize{http://www.seeedstudio.com/wiki/Category:Grove}}. 

\subsection{Smart Lighting Operations}
\label{operation}
Our solution relies on three main operations include both monitoring and controlling as summarized below.

	\subsubsection{Data dissemination from luminosity detectors}
	
Each luminosity detector measures the light level periodically and pushes (sends) Interest notifications that include the light level value and the time stamp to the smart controller. To receive such notifications, the controller has to register the name prefix with the home router through the static routing.  As the luminosity detector frequently measures and disseminates data periodically, the acknowledgement is omitted to avoid overhead messages.

	\subsubsection{Data dissemination from occupancy detectors}

	Occupancy detectors are used to detect users' movements (i.e, IN or OUT). Unlike luminosity detectors, movement data is crucial for controlling the light (i.e., switching ON or OFF), the acknowledgements are expected from the smart home controller to ensure the reliable message transmission. The occupancy detectors can retransmit the same message unless acknowledgements are received within a desired interval.
	\begin{figure}[!ht]
  \centering
    \includegraphics[width=0.4\textwidth]{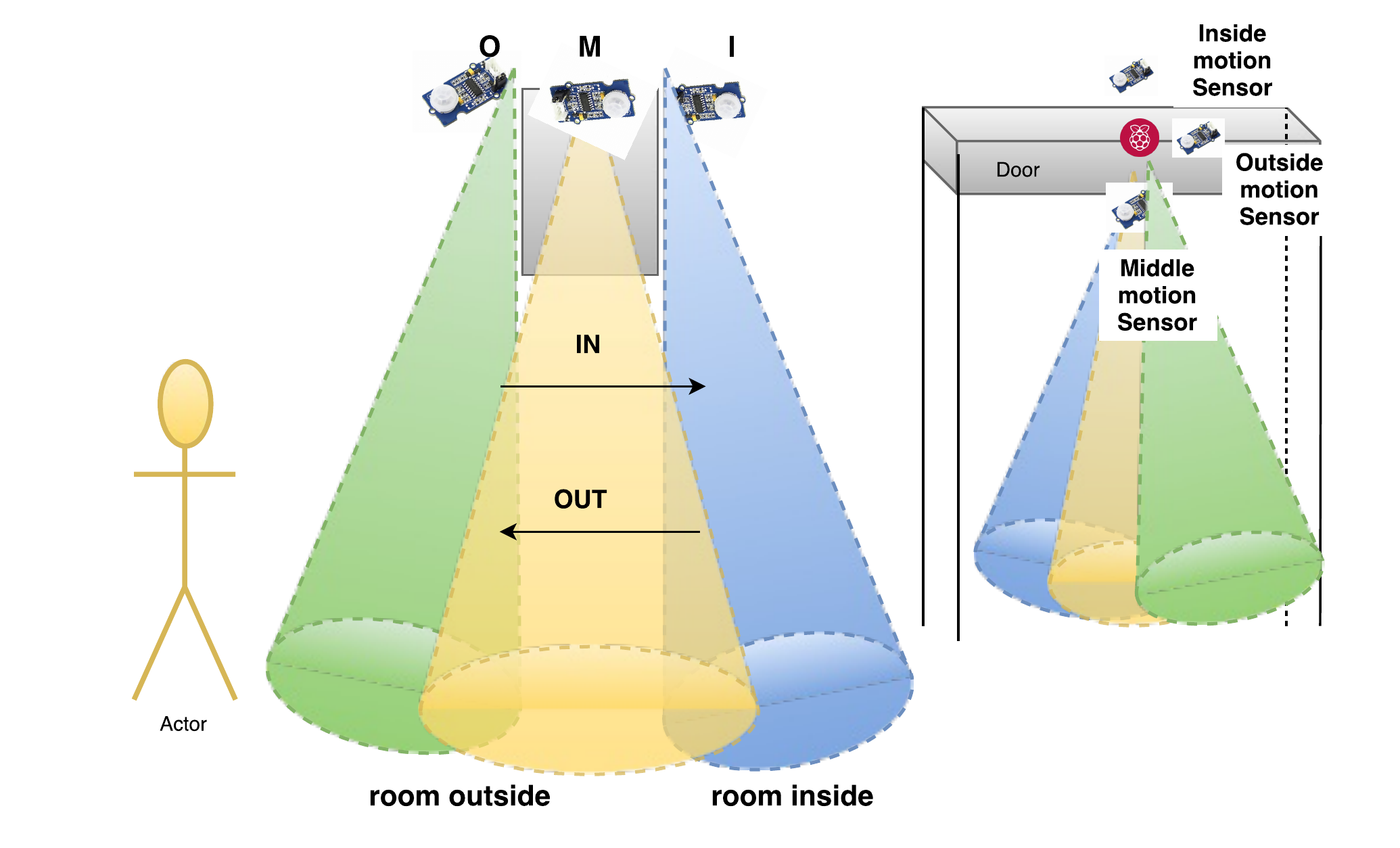}
    \vspace{-1.5ex}
     \caption{Occupancy Detector implementation}
      \label{occ_detector}
 \end{figure}

We deploy three motion sensors to derive the direction of users' movements based on the pattern of motion detection over time. Three sensors are identified as $O$, $M$ and $I$ which are placed outside, middle and inside of the door frame respectively (see Figure~\ref{occ_detector}). 

Based on the motion detection of $OMI$, the occupancy detector deduced the direction of movement. If $OMI$ status changes from 110 to 011, it is considered as IN movement while $OMI$ status change from 011 to 110 is considered as an OUT movement. 
If the transition is not occurred within a pre-defined delay period, the detector resets IN, OUT waiting flags and refreshes detecting process. This delay period is used to rule out possible interferences from users' movement only inside or outside the room. We enforce the sensors to have narrow beam width to mitigate these interferences. 

	\subsubsection{Light Control}
	The smart home controller continuously collects data from luminosity detectors and occupancy detectors. The control commands are issued based on users' movements and light level. The room light level is compared with specific $Min$ and $Max$ thresholds, separately defined for different functional areas such as bed rooms and kitchens.

	\section{Applying NDN to Smart Lighting Solution}
In this section, we describe the potential functions adapted from primitive NDN to support smart lighting's operations. 

\subsection{Naming Scheme}

\begin{figure}[!ht]
  		\centering
    		\includegraphics[width=0.45\textwidth]{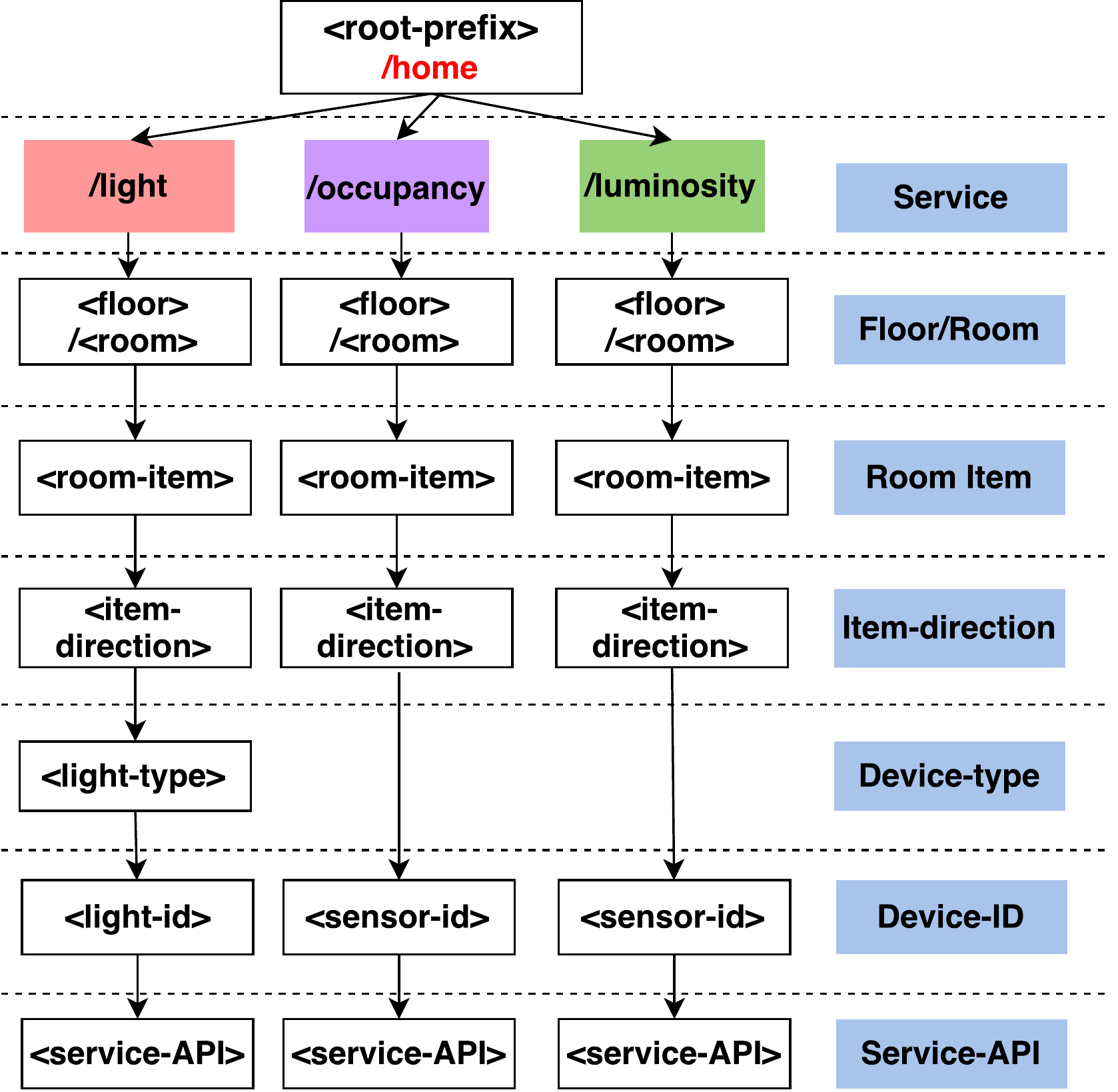}
    		\vspace{-1.5ex}
     	\caption{Naming scheme}
      \label{naming}
            \vspace{-2ex}
	\end{figure}

NDN supports semantic names at the network layer, makes proper naming design a high priority when creating the application. We define a hierarchical naming scheme for the solution based on services semantic and physical layout of the floor plan as presented in Figure~\ref{naming}. At the top level, a root prefix is defined as \emph{/home}, following service name components. In the lighting solution, there are three main services: light, luminosity and occupancy. The naming scheme can be expanded as more services are deployed. Next, we define the hierarchical location components such as floor, room, room items (i.e., wall, ceiling, table) and item orientation. When different types of devices are available within the same service, the device type can potentially be appended, based on a unique device ID which can be a unique integer. Finally, name components that represent the service interface of each device can be appended, including methods and attributes supported by each device. For example, light nodes may support switchON, switchOFF methods while sensors may support read method.

\subsection{Smart Lighting Communication Models}
NDN inherently supports the pull-based communication. However, in applications like our smart lighting, the push-based data dissemination is more appropriate. For instance, the occupancy detector can push data whenever a movement is detected. 
	\subsubsection{Push/Publish based data dissemination}
There are two possible approaches to implement the push/publish based communication using NDN. First, if the data's format is text (e.g., short string characters), the data can be directly sent by appending it to the Interest name. This method is commonly referred to as Interest notification~\cite{amadeo:ICC_home}. 
Second, if the data is embedded as a file (e.g., .txt, .zip), the data producer can send an Interest to the consumer which in turn triggers an Interest from the consumer to fetch data~\cite{Burke:NCA-2014}. In this case, if data is cached, it can induce message overhead and latency. In our solution, we implemented the first option for sensor data dissemination, since data is generated in string format. 

\begin{figure}[!ht]
  		\centering
    		\includegraphics[width=0.4\textwidth]{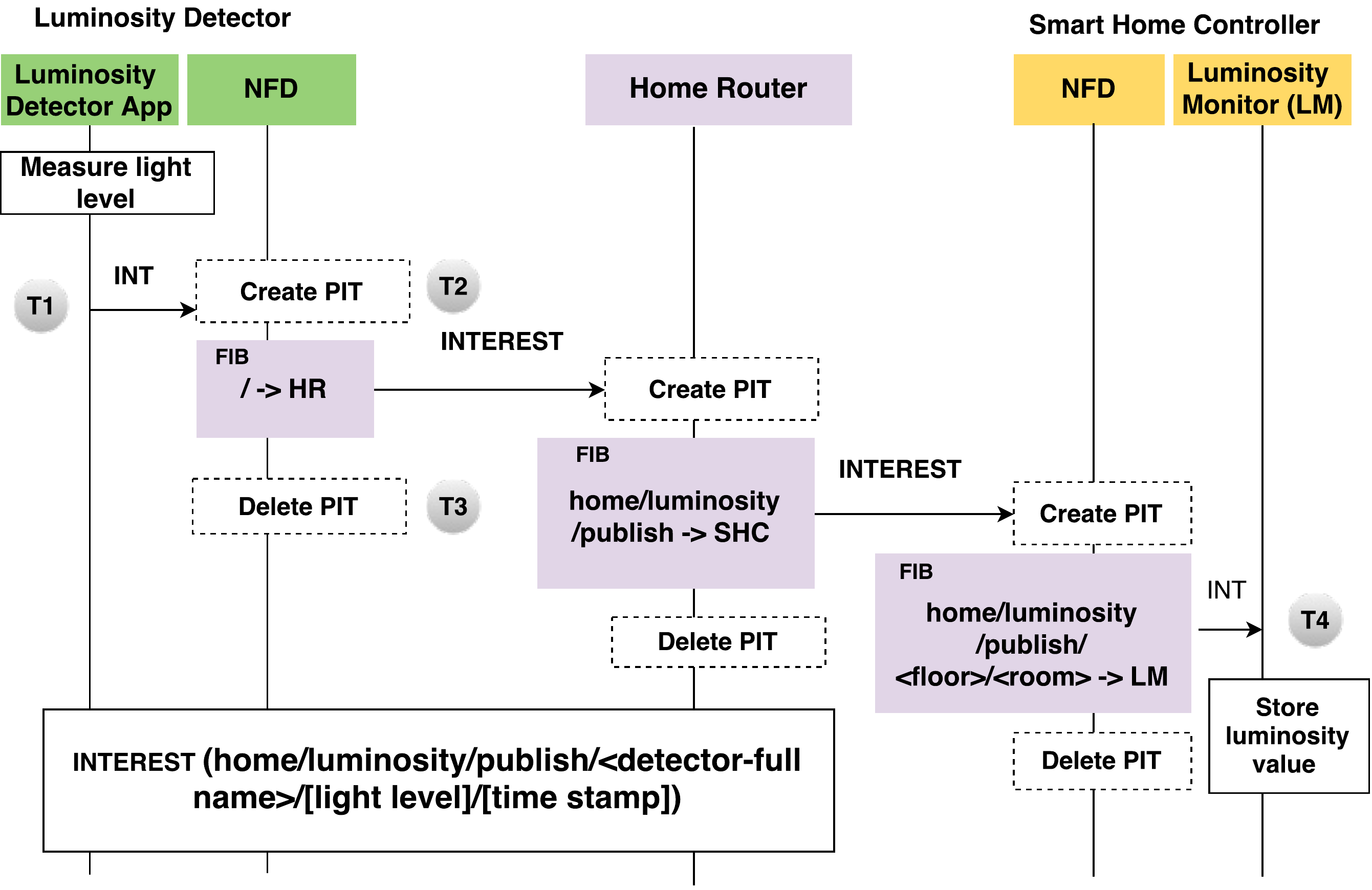}
    		\vspace{-1.5ex}
     	\caption{Data dissemination from luminosity detectors via push based communication}
      \label{luminosity_msg}
	\end{figure}

Figure~\ref{luminosity_msg} illustrates the message flow between luminosity detectors and luminosity monitor via the home router. 

The luminosity detector is configured to measure light levels every five seconds and to push Interest notification by appending the light level and the time stamp to the name. We distinguish between Interest notification name and regular name by adding the fixed name component \emph{publish}€™ after the service name component in the name hierarchy. Consequently both push and pull based data disseminations can be implemented on the same sensor. 

To subscribe the Interest notifications, the particular name prefixes (e.g., \emph{/home/luminosity/publish}) must be registered in the FIB of smart home controller.  

Corresponding PIT entries are deleted after the Interest life time (T3-T2) which is set to two seconds by the detector application. Specifically, small Interest life time is sufficient to maintain PIT records as there is no Data message in return. Setting  Interest life time to zero can avoid creating PIT entries, but may cause Interest loops depending on routing entries. 

\begin{figure}[!ht]
  		\centering
    		\includegraphics[width=0.4\textwidth]{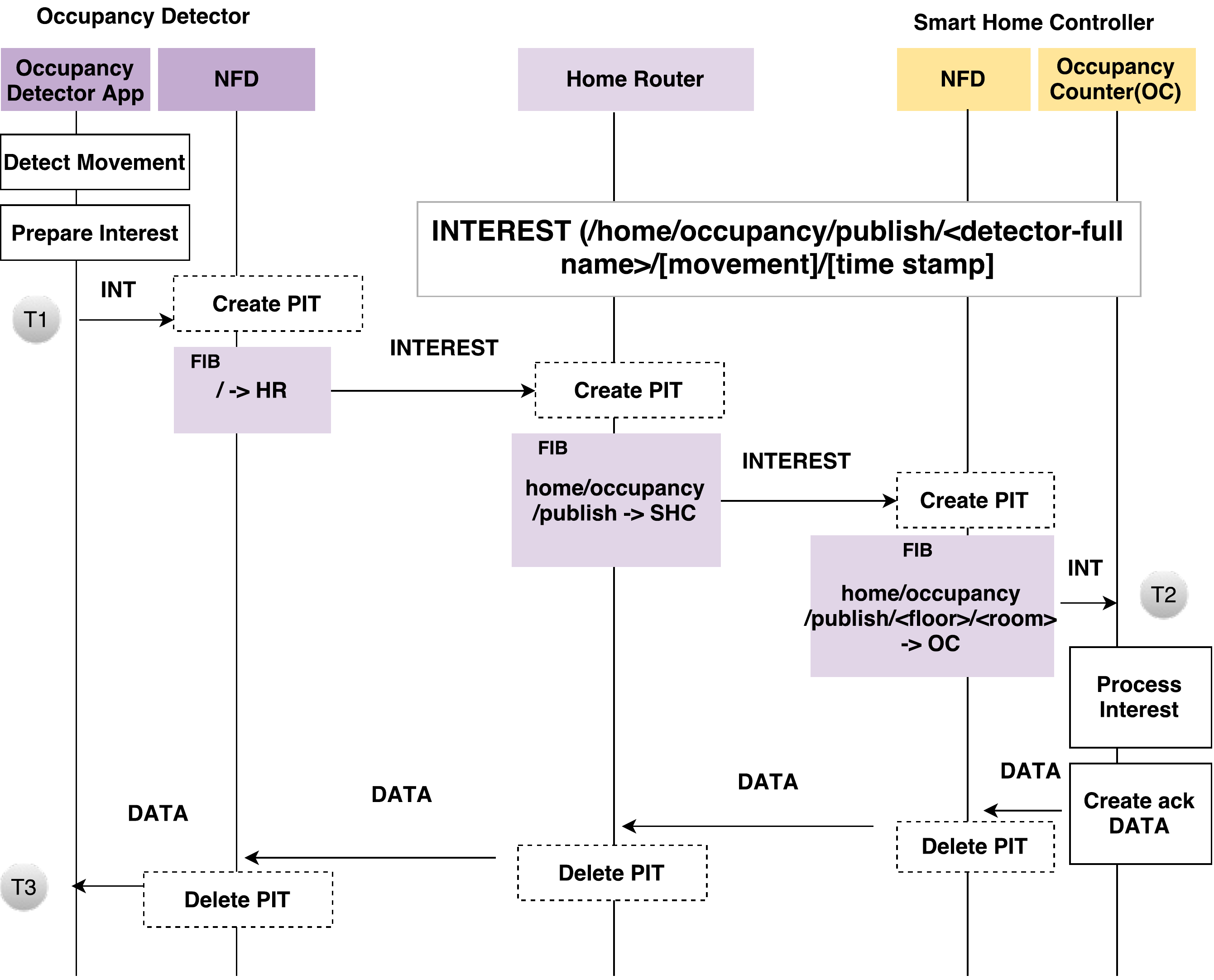}
    		\vspace{-1.5ex}
     	\caption{Controlling lights via pull based communication}
      \label{occupancy_msg}
	\end{figure}

Figure~\ref{occupancy_msg} illustrates Interest/Data exchange between occupancy detectors and occupancy counter via the home router.  The occupancy detector pushes an Interest message appending the movement data into the notification NDN name. The occupancy counter in smart home controller registers the name prefix \emph{/home/occupancy/publish/} in order to continuously receive the occupancy notification Interests. Here, we use dummy Data message as the acknowledgement to confirm the Interest reception. 

Currently, Interest life time is set to four seconds and each Interest message is retransmitted up to three times in case of time-outs. 
	\subsubsection{Pull based controlling}
\begin{figure}[!ht]
  		\centering
    		\includegraphics[width=0.4\textwidth]{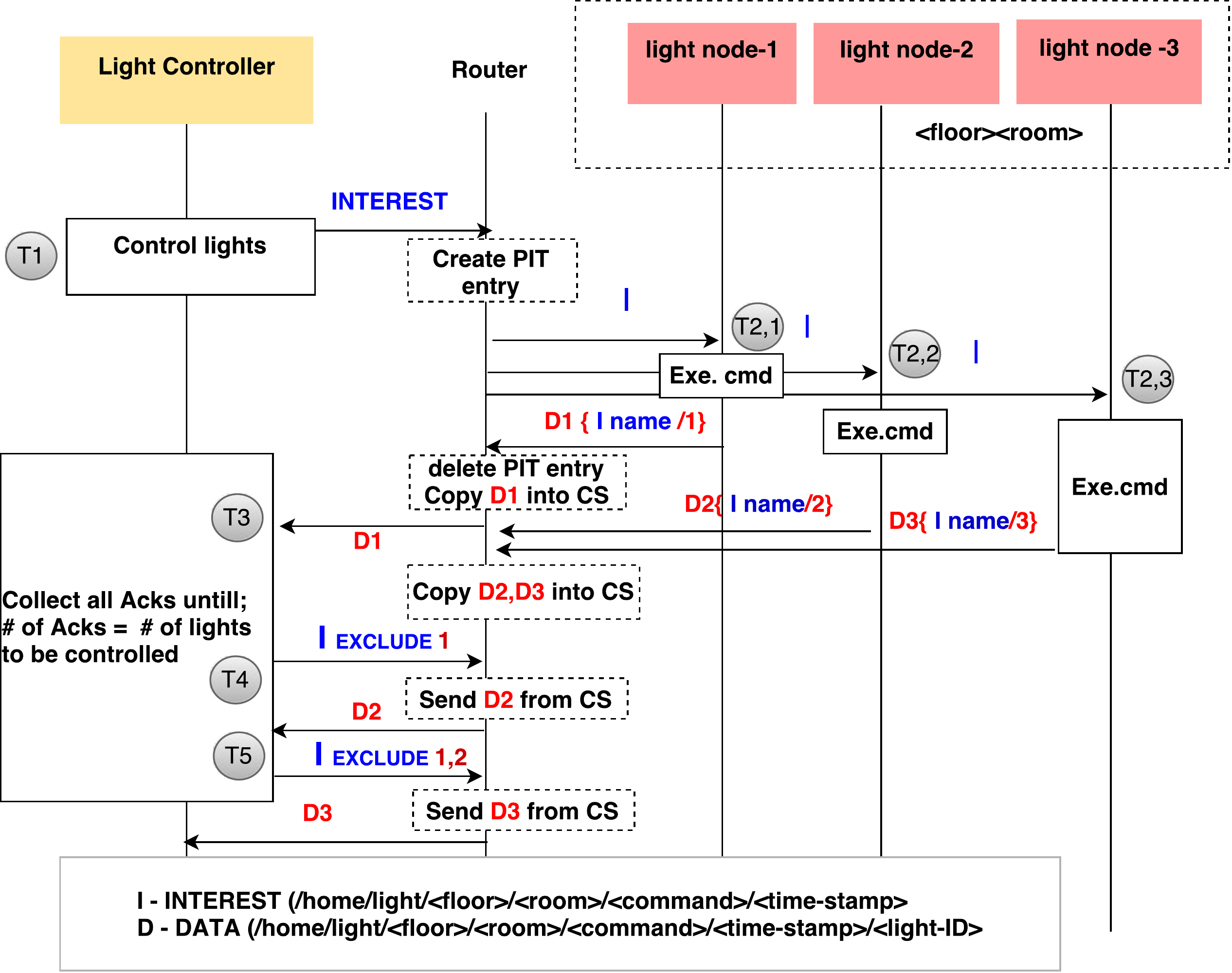}
    		\vspace{-1.5ex}
     	\caption{Controlling lights via pull based communication}
      \label{light_msg}
	\end{figure}

To control devices, we apply NDN default pull based communication in our scenario by appending commands into Interest name. Upon executing the command, actuators reply with ACK Data. In addition, authenticate Interests can be used to embed controller identification into the Interest message by appending controller signature. Figure~\ref{light_msg} illustrates the controlling lights using pull based communication. Smart home controller uses NDN inherent multicast forwarding feature to control group of lights simultaneously. 

In the solution, we used static routing in the home router. Table~\ref{table:model} summarizes the communication models implemented in our system.
\vspace{-2ex}

\begin{table}[ht]\scriptsize
\caption{Communication models for smart home lighting system}
\centering
\begin{tabular}{|l|l|l|l|l|}
\hline
\textbf{Operation}                                                                     & \textbf{Model}                                                   & \textbf{Nature}                                        & \textbf{\begin{tabular}[c]{@{}l@{}}Producer\end{tabular}} & \textbf{\begin{tabular}[c]{@{}l@{}}Consumer\end{tabular}} \\ \hline
\begin{tabular}[c]{@{}l@{}}Data dissemination \\ from LD\end{tabular}                  & \begin{tabular}[c]{@{}l@{}}Interest \\ notification\end{tabular} & \begin{tabular}[c]{@{}l@{}}Many \\ to One\end{tabular} & LD                                                           & SHC                                                          \\ \hline
\begin{tabular}[c]{@{}l@{}}Data dissemination \\ from OD\end{tabular}                  & \begin{tabular}[c]{@{}l@{}}Interest \\ notification\end{tabular} & \begin{tabular}[c]{@{}l@{}}Many \\ to One\end{tabular} & OD                                                           & SHC                                                          \\ \hline
\begin{tabular}[c]{@{}l@{}}Light Controller \\ issues control \\ messages\end{tabular} & \begin{tabular}[c]{@{}l@{}}Pull based \\ controling\end{tabular}       & \begin{tabular}[c]{@{}l@{}}One \\ to Many\end{tabular} & LN                                                          & SHC                                                          \\ \hline
\end{tabular}
  \label{table:model}
\end{table}

\subsection{Unicast and Multicast via Name}
Light controller module in SHC is responsible for sending commands to light nodes in the house. The transmission is supported by both unicast (e.g., an Interest command is sent to each particular light node) and multicast (i.e., a single Interest command can be sent to a group of light nodes). Given that the number of lights to be controlled is $N$;

	\textbf{\emph{Unicast control:}} $N$ number of individual Interests are sent in the format of \emph{/home/light/$<$floor$>$/$<$room$>$/$<$light-ID$>$/$<$command$>$}. 
Each light node has to register its name prefix in the home router with the aforementioned format causing $N$ entries in the FIB. Due to sequential Interests, light operation latency may also increase for subsequent lights.
	
\textbf{\emph{Multicast control:}} Our solution benefits from the inherent multicast communication of NDN by using an aggregated namespace. A single Interest packet is enough to command tasks to multiple devices. As illustrated in Figure~\ref{light_msg}, all light nodes in floor1 register a common prefix \emph{/home/light/floor1/} in their local forwarder. After receiving the Interest message command (T2.1, T2.2 and T2.3), each light node filters matching Interests and simultaneously executes the command (switch off the light). To ensure the command execution, each light node is required to reply with ACK Data by appending own light ID to the received Interest name. Multiple matching ACK Data messages are expected to be sent to the home router, but only the first matching ACK Data is forwarded back to the controller, due to the self traffic regulation feature of NDN. However both solicited and unsolicited Data are stored in the router cache by default~\cite{NFD_developper}. 
As a consequence, the light controller resends the same Interest, setting exclude filter with light IDs in previously received acknowledgements and fetches remaining ACK Data one by one from router cache until it fetches all the ACKs equal to number of lights to be controlled. 

Data in cache becomes stale after the time period set in \textit{DataFreshnessPeriod} 
and Interests having \textit{MustBeRefresh} flag are not replied back with the stale Data. In lighting application, command Interests are sent with \textit{MustBeFresh} flag to fetch fresh acknowledgements and to track light operation in real time. 
Accordingly, sum of RTT of fetching ACKs should be smaller than \textit{DataFreshnessPeriod}. If controller could not fetch the acknowledgement from the router cache, the command Interest is rebroadcasted causing message overhead.
Intuitively, the multicast control with filtering method provides the flexibility to group multiple nodes with a single FIB record. 
With the conventional IP approach, IP multicast groups have to be created for each such unique group required. However, with NDN, using the semantic of name components, light nodes could be grouped meaningfully in addition to namespace aggregation in the name hierarchy.

\section{Performance Evaluation}
We implemented the NDN smart home lighting testbed in our laboratory while deploying three home nodes with light bulbs, light sensors and motion sensors (see Figure~\ref{architecture}). The users' movements were simulated to avoid the impact of possible random occupancy detection failures of the real occupancy detector and to have a uniform movement detection pattern. Therefore, we simulated a sequence of alternative IN, OUT movements for every 1 minute while considering $20$ movements in total. During the experiments, we set up a local home WiFi network isolated from the Internet.

We compared our system with NetPie, a local cloud platform\footnote{\scriptsize{NetPie is one of the most popular IoT cloud platforms in Thailand, currently running more than 1500 applications, connecting around 14000 things. (https://netpie.io)}} with its own instant messaging system that supports the MQTT protocol and the HTTP RESTful API. 
To connect our platform to NetPie, we implemented client modules for each device and registered them with the NetPie server. We configured the modules with the matching identities and application id (home), so that they can authenticate and publish the data under URI-like topic names.

\begin{figure}[!ht]
  \centering
    \includegraphics[width=0.45\textwidth]{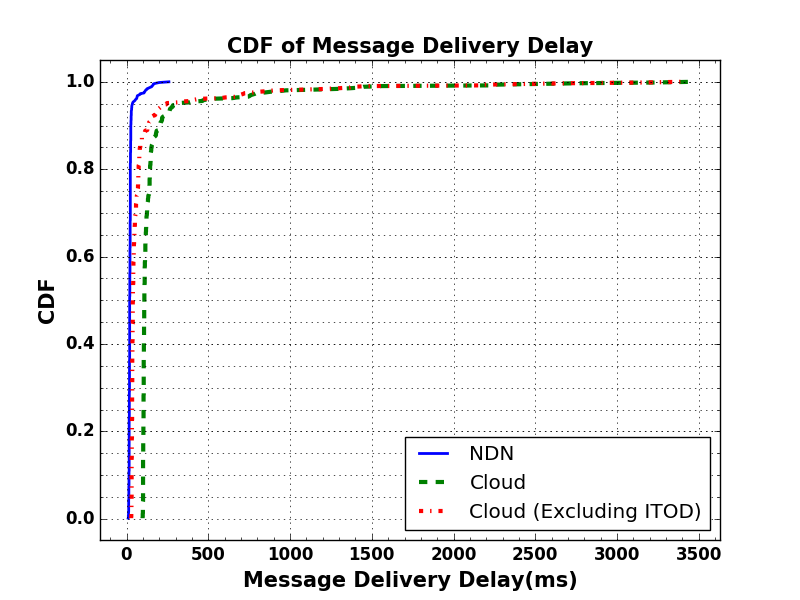}
    \vspace{-1.5ex}
     \caption{CDF of message delivery delay}
      \label{latency_cloud}
\end{figure}

We base our comparative evaluation with message delivery delay which is the time taken to deliver a message from the producer to consumer (see Table~\ref{table:model}). 
Figure~\ref{latency_cloud} shows the CDF of message delivery delay for three operations under NDN and cloud approaches. With the cloud, we separately measured Internet Transmission Overhead Delay (ITOD) by pinging the NetPie server from the home router and estimated it to be approximately $70$ ms. We deducted this overhead to model a locally operated could.
The plot shows that $95\%$ of messages have been transmitted with delays of less than 36 ms and 315 ms in the NDN and cloud respectively. When Internet transmission overhead is deducted from the cloud approach, $95\%$ of messages have been transmitted in less than 245 ms. This indicates that even if the cloud is hosted locally, the delay is still higher than in NDN due to processing overhead in the cloud platform and its protocol stack. In addition, the maximum delay in the cloud and cloud without overhead are increased up to 3.46 and 3.39 second while maximum delay of NDN is only 0.26 second. In addition, under NDN, all the messages have been delivered with delays less than 500 ms which is the recommended value specified in RFC 5826 ~\cite{RFC_5826} for home automation applications.
Delay reduction is crucial for the applications like smart lighting, since it can guarantee real time response and simultaneous controlling sets of lights.

In both NDN and cloud based approaches, light nodes are operated from a single Interest, however, due to self traffic regulation feature of NDN, (N-1) extra Interests are sent from the controller to the router to fetch all ACK Data. These subsequent Interests include an extra controlling field, an exclude filter which adds additional bytes ( 2 controlling bytes and 3 more bytes for each excluding light ID) to the Interest header. 

\vspace{-1.5ex}
\section{Analysis on FIB Scalability}

\begin{figure}[!ht]
  \centering
    \includegraphics[width=0.4\textwidth]{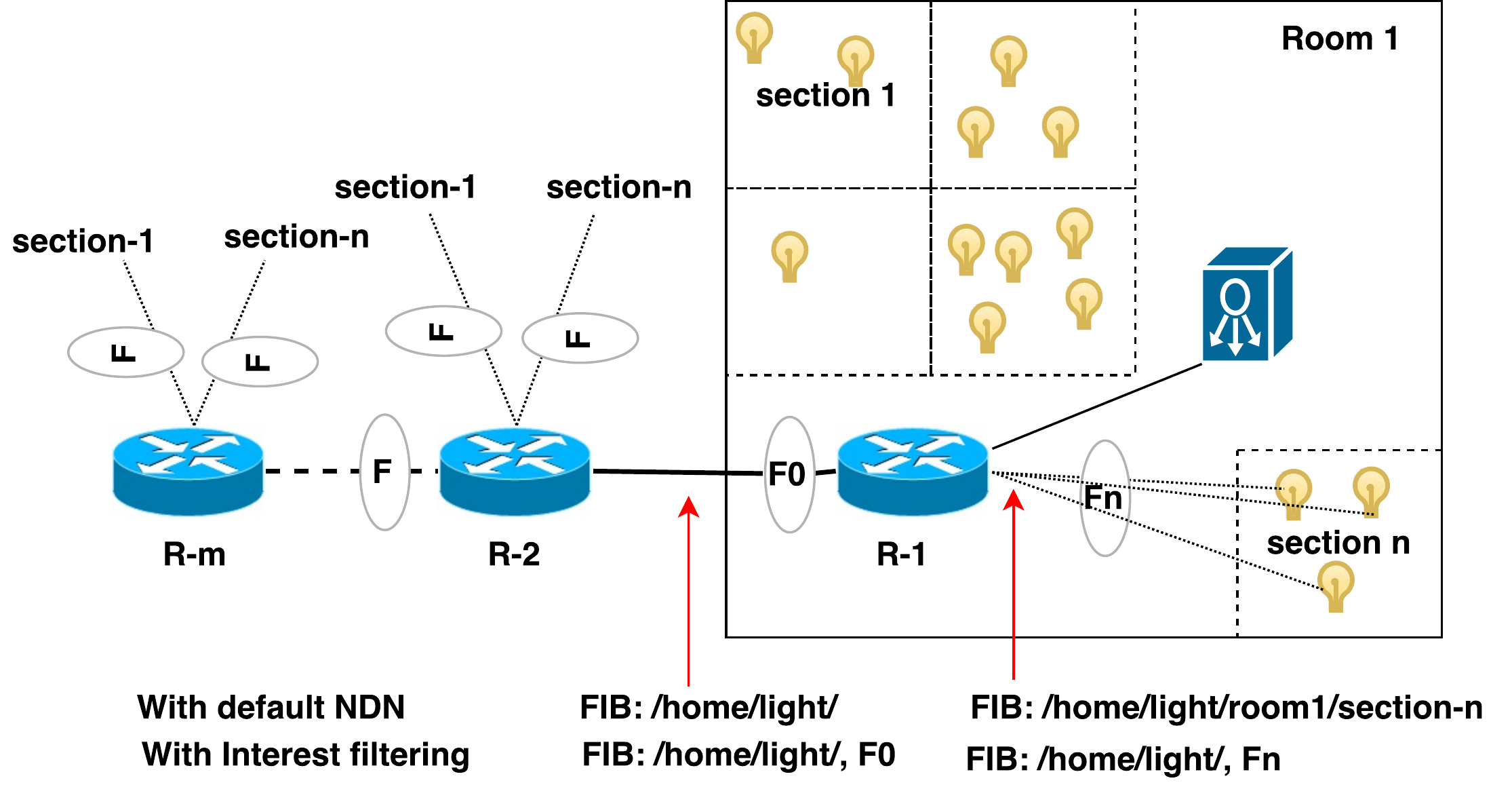}
    \vspace{-1.5ex}
     \caption{Sample topology for FIB scalability analysis}
      \label{FIB scalability}
          \vspace{-1.5ex}
\end{figure}

Since we have designed a solution, which focuses control lights in particular areas ( e.g., rooms in homes), we analyze the scalability when it is deployed in the larger buildings with many controlling areas like an exhibition hall. We consider a linear topology as in Figure~\ref{FIB scalability}. We assume exhibition area covers with $m$ number of linearly connected routers, each router serving a single room. Smart home controller is connected to a corner router which is assumed to be router 1. Each room has an arbitrary number of sections $n$, thus each router must include minimum number of routing entries in FIB which is equal to $n+1$.

%\vspace{-2ex}

We claim that if router can set Interest filters towards faces (e.g., $F0$, $Fn$ in Figure~\ref{FIB scalability}), it is possible to set one FIB entry in each router and set appropriate filters towards each face.
For example, router 1 can have filter as Table 2. Therefore with proper Interest filters, each router can reduce number of routing entries in their FIB from $n$+1 to 1. With current fixed linear topology, $m$ does not impact the number of routing entries per router, but if the topology changes, number of routing entries increases with $m$. For example, in mesh topology or if smart home controller is moved to another router in this linear topology, more routing entries are required to segregate traffic to reduce unneccessary flooding. With our suggestion of Interest filtering, a single FIB entry can serve for all routing requirements, given that filters are set properly over the router interlinks.
\vspace{-1ex}
\begin{table}[ht]\scriptsize
\caption{FIB entry and filtering of Router 1 in exhibition hall scenario}
\centering
\label{table:filter}
\begin{tabular}{|l|l|l|}
\hline
\multicolumn{1}{|c|}{\textbf{FIB entry}} & \multicolumn{1}{c|}{\textbf{Face}}                                          & \multicolumn{1}{c|}{\textbf{Filter}}                                                                                                                               \\ \hline
\multicolumn{1}{|l|}{/home/light}       & \begin{tabular}[c]{@{}l@{}}To Router 2\\ (room 2 to m)\end{tabular}        & \begin{tabular}[c]{@{}l@{}}Filter 0 (F0):{[}\textless room-2\textgreater ...\textless room-m\textgreater{]}\textless\textgreater*\end{tabular}              \\ \hline
\multicolumn{1}{|l|}{}                  & \begin{tabular}[c]{@{}l@{}}To all lights in \\ room1 section-1\end{tabular} & \begin{tabular}[c]{@{}l@{}}Filter 1 (F1):{[}\textless room-1\textgreater{]}{[}\textless section-1\textgreater{]}\textless\textgreater*\end{tabular} \\ \hline
\multicolumn{1}{|l|}{}                  & \begin{tabular}[c]{@{}l@{}}To all lights in \\ room1 section-n\end{tabular} & \begin{tabular}[c]{@{}l@{}}Filter n (Fn):{[}\textless room-1\textgreater{]} {[}\textless section-n\textgreater{]}\textless\textgreater*\end{tabular} \\ \hline
\end{tabular}

\end{table}	

\vspace{-1ex}
\section{Discussion}

Based on our experience in developing smart lighting with NDN, we envision that there is still a lot of room to improve NDN to develop sustainable IoT applications. Effective push based and multi-source communication are especially viable in most IoT scenarios while IoT routing protocols can enable automatic service publishing in the larger networks. Implementing push model on top of NDN as Interest notification is doable, but it has the drawback of inability of caching contents. The fundamental publish-subscribe model should be considered for NDN as a subscriber which can automatically receive data whenever a producer publishes a new content. With Interest filtering, we could enable lights to provide service on multiple meaningful names with a single FIB entry. However, current Interest filtering is only supported by NDN application client. Implementing the same feature in the core NDN router will enable customization in FIB scaling with respect to node resource availability. Besides, Interest filtering introduces implicit complexity in terms of analysing the incoming Interest message which may effect to the processing delay.

Multicast via name is naturally supported by primitive NDN. 
However, application has to fetch each acknowledgment with repetitive Interests causing significant message overhead. In this context, some techniques like long lived Interest concept~\cite{carzaniga:pub_sub} and multi-source communication~\cite{marica:NDN_multi} could be explored to develop the efficient communication protocol where multiple Data can be fetched with a single Interest.

\section{Conclusion and Future work}
   We explored the flexibility of NDN over two main aspects of IoT applications, including data dissemination and command execution with a prototype implementation.

NDN enables network entities to be identified with human readable English names based on home semantics, keeping consistent naming in both application and network layer. To enable periodic and event based data dissemination with a lower message overhead and latency, our solution applies push based communication through the Interest notification. 
We use multicast control of lights with Interest filtering in the application layer for effective group communication. 

We are currently in the process of deploying our NDN based smart light solution in the wild - at a residence within the university following a comprehensive evaluation on both technical and user experiences.
Meanwhile, we plan to explore long lived Interest concept and multi-source communication to enhance the networking performance of the solution.
Enabling NDN security features effectively with a suitable trust model and light weight security algorithms is also in our roadmap to ensure authenticated light control.

%\bibliographystyle{IEEEtran}
%\bibliography{biblio_lighting}

\end{document}